# From Public Square to Echo Chamber: The Fragmentation of Online Discourse

Abhinav Pratap
Department of Computer Science
and Engineering, ASET
Amity University, Noida, India
TheAPratap@gmail.com

Amit Pathak
Department of Computer Science
and Engineering, ASET
Amity University, Noida, India
pathakap364056@gmail.com

*Abstract*—This paper examines how social media algorithms and filter bubbles contribute to the fragmentation of online discourse, fostering ideological divides and undermining shared understanding. Drawing on Michael Sandel's philosophical emphasis on community and shared values, the study explores how digital platforms amplify discrimination discourse—including sexism, racism, xenophobia, ableism, homophobia, and religious intolerance—during periods of heightened societal tension. By analyzing the dynamics of digital communities, the research highlights mechanisms driving the emergence and evolution of discourse fragments in response to real-world events. The findings reveal how social media structures exacerbate polarization, restrict cross-group dialogue, and erode the collective reasoning essential for a just society. This study situates philosophical perspectives within a computational analysis of social media interactions, offering a nuanced understanding of the challenges posed by fragmented discourse in the digital age.

## I. INTRODUCTION

The rise of social media has profoundly transformed public discourse, redefining the concept of community in the digital age. While these platforms enable unprecedented connectivity, they also fragment communication into isolated and polarized subgroups. Social media algorithms, designed to maximize engagement, often prioritize sensational and divisive content, fostering echo chambers that limit exposure to diverse perspectives [1]. This phenomenon amplifies discrimination discourse—including sexism, racism, xenophobia, ableism, homophobia, and religious intolerance—leading to the erosion of shared understanding and weakening the collective reasoning necessary for a cohesive society [2].

Michael Sandel's philosophical exploration of justice and community underscores the importance of shared values and mutual accountability in sustaining a just and equitable society [3]. He contends that community is not merely a network of individuals but a foundation for moral and civic engagement. However, the mechanisms underpinning social media platforms frequently undermine these principles. By tailoring content to user preferences and promoting divisive narratives, these platforms contribute to the fragmentation of public discourse, isolating users in self-reinforcing silos and amplifying ideological divides [4].

This study examines how social media platforms shape the dynamics of online communities, focusing on the fragmentation of discourse in response to real-world events. Discrimination discourse, which emerges prominently during moments of societal tension, serves as a lens for analyzing the interplay between algorithmic curation, user behavior, and societal polarization [5].

By situating these findings within broader philosophical and sociological frameworks, this study seeks to illuminate the mechanisms driving the fragmentation of online discourse and its impact on the concept of community in a digital age.

## II. LITERATURE REVIEW

The fragmentation of online discourse has emerged as a critical concern in the digital age, characterized by the splintering of conversations into isolated subgroups that undermine collective reasoning. This phenomenon has attracted significant scholarly attention across disciplines, encompassing communication studies, sociology, political science, and philosophy. This review synthesizes key insights, focusing on the role of social media algorithms, the amplification of discrimination discourse, and the implications for community cohesion in light of Michael Sandel's emphasis on shared values and collective accountability.

### A. Redefining Community in the Digital Age

The concept of community has undergone profound transformation with the advent of digital platforms. Traditionally rooted in geographic proximity and shared cultural or social values [6], community has become increasingly virtual, transcending spatial boundaries. Social media platforms enable individuals to form communities around shared interests, often empowering marginalized voices [7]. For instance, online platforms have provided spaces for LGBTQ+ individuals and people with disabilities to build networks of support and visibility.

However, the very features that foster inclusivity can also contribute to exclusivity. Scholars have highlighted the dual nature of online communities, where anonymity and algorithmic curation facilitate the proliferation of extremist ideologies [8]. Alt-right groups, for instance, have exploited these dynamics to disseminate polarizing rhetoric and normalize hate speech [9]. This duality underscores the challenge of sustaining cohesive communities in an environment dominated by personalized content delivery systems.

*B. Discourse Fragmentation: Causes and Consequences*

Discourse fragmentation refers to the division of public conversation into isolated sub-discourses that inhibit meaningful dialogue. Social media platforms, driven by algorithms designed to maximize user engagement, play a central role in this process. These algorithms prioritize content that elicits strong emotional responses, often amplifying divisive narratives and reinforcing user biases [10].

Discrimination discourse—encompassing sexism, racism, xenophobia, ableism, homophobia, and religious intolerance—illustrates the mechanisms and impacts of fragmentation. Dezhboro et al. (2024) identified these fragments as dominant themes during moments of societal tension, such as the COVID-19 pandemic and the Black Lives Matter movement [11]. Their findings revealed how real-world events catalyze the rapid emergence and dissolution of fragmented communities, further intensifying polarization.

The consequences of fragmentation extend beyond online spaces. Research shows that ideological silos restrict exposure to diverse perspectives, entrenching users within echo chambers that reinforce existing beliefs. This dynamic not only exacerbates polarization but also weakens the collective ability to address complex societal challenges, as discourse becomes increasingly fragmented and adversarial.

*C. The Role of Algorithms in Shaping Discourse*

Social media algorithms disproportionately amplify polarizing content, creating a feedback loop that reinforces ideological divides [12]. Studies have shown that algorithms selectively expose users to content that aligns with their preferences, limiting opportunities for cross-group dialogue [13].

This algorithmic bias is particularly pronounced in the context of discrimination discourse. Research has demonstrated that hateful content often gains disproportionate visibility due to its high engagement metrics, perpetuating narratives of exclusion and marginalization. During real-world crises, such as pandemics or social justice movements, these dynamics are further exacerbated, as public anxieties and tensions become fertile ground for divisive rhetoric.

*D. Philosophical Perspectives on Fragmentation*

Michael Sandel argues that a just society requires a shared understanding of the common good, rooted in community and collective reasoning [3]. However, the individualistic ethos promoted by digital platforms undermines these ideals, prioritizing personal preference and engagement over shared accountability [14].

From a sociological perspective, the decline of community-based social capital further exacerbates the fragmentation of discourse. Putnam (2000) observes that the erosion of traditional, place-based communities has diminished networks of trust and reciprocity, making it increasingly difficult to sustain cohesive public discourse. This fragmentation is reflected in the dynamics of online interactions, where ideological divides are amplified by the lack of accountability and the transient nature of digital communication.

*E. Real-World Events and Their Impact on Online Discourse*

Real-world events play a critical role in shaping the dynamics of online discourse. Moments of societal tension—such as the COVID-19 pandemic, the Black Lives Matter movement, and contentious elections—serve as catalysts for the amplification of discrimination discourse. Dezhboro et al. (2025) demonstrated how these events influence the formation and evolution of fragmented communities, revealing the interplay between real-world circumstances and digital interactions.

For example, during the COVID-19 pandemic, xenophobic rhetoric targeting Asian communities surged, fueled by algorithmic amplification and political narratives (United States Commission on Civil Rights, 2023). Similarly, the Black Lives Matter movement highlighted how social media platforms amplify both advocacy and opposition, creating fragmented spaces for discourse that often devolve into polarization. These examples underscore the complex relationship between societal events and the structural dynamics of digital platforms.

*F. Gaps in Existing Research*

Despite extensive scholarship on discourse fragmentation and online communities, significant gaps remain. Much of the existing literature focuses on individual platforms or specific events, limiting the generalizability of findings. Moreover, the interplay between algorithmic curation and user behavior in fostering fragmentation is not fully understood, particularly in the context of discrimination discourse.

Another critical gap lies in the exploration of solutions for fostering cohesive public discourse in fragmented digital spaces. While scholars have highlighted the role of algorithmic bias in shaping discourse, there is limited research on strategies for mitigating these biases and promoting cross-group engagement. Addressing these gaps is essential for advancing the understanding of discourse fragmentation and its implications for society.

III. METHODOLOGY

This research employs a multi-layered methodological framework to analyze the fragmentation of online discourse and its impact on community formation. By integrating computational and qualitative approaches, the study investigates how discrimination discourse—comprising sexism, racism, xenophobia, ableism, homophobia, and religious intolerance—emerges and evolves on social media platforms, particularly during moments of societal tension.

*A. Research Design*

The study adopts a mixed-methods approach, combining social network analysis with thematic and temporal analyses. This approach enables a comprehensive exploration of both the structural and thematic dimensions of discourse fragmentation, focusing on:
- The dynamics of community formation and dissolution.

- The influence of real-world events on discourse dominance.
- The amplification of discrimination discourse and its fragments in online interactions.

*B. Data Sources*

The primary data consists of publicly available social media content, focusing on platforms such as Twitter, which are widely used for public discourse. The data selection criteria include:
- Content Focus: Posts containing keywords and hashtags related to discrimination discourse and significant real-world events (e.g., COVID-19 pandemic, Black Lives Matter movement, or elections).
- Interaction Data: User interactions such as retweets, replies, and mentions that reflect the structure of online networks.
- Temporal Scope: Data collected during and around major societal events to capture the emergence and evolution of discourse fragments.

*C. Data Collection*

Data was collected using an automated process that adhered to the following steps:
- Keyword and Hashtag Filtering: Social media posts were extracted using predefined keywords and hashtags representing discrimination discourse (e.g., #BlackLivesMatter, #StopAsianHate).
- Time-Bound Sampling: Posts were gathered over a defined timeframe to capture trends and shifts in discourse. For example, data was collected before, during, and after significant events to observe their influence on online interactions.
- Data Preprocessing: The raw data underwent preprocessing to ensure quality and relevance. Steps included removing duplicates, eliminating spam, normalizing text (e.g., lemmatization, stopword removal), and retaining posts with significant engagement (e.g., likes, retweets).

*D. Data Analysis*

I. Thematic Analysis

The textual content of the posts was analyzed to identify dominant themes and patterns within the discourse. Using natural language processing (NLP) techniques, the study categorized posts into specific fragments of discrimination discourse, including sexism, racism, xenophobia, ableism, homophobia, and religious intolerance.

Key aspects of thematic analysis included:
- Content Focus: Identifying the primary topics driving engagement within the discourse.
- Polarization: Examining how divergent views on these topics contributed to ideological silos.
- Ephemerality: Observing how long topics persisted and the speed at which new themes emerged.

TABLE I. USERS ENGAGING IN HATE SPEECH TWEETS

| City | Tweets | Users | Multi-hate users |
|---|---|---|---|
| NYC | 1,139,236 | 457,503 | 105,904 |
| SF | 1,305,772 | 567,557 | 131,700 |
| Seattle | 1,286,526 | 530,365 | 128,086 |

To support this, Table I illustrates a representative distribution of users engaging in various forms of discriminatory language across cities. This distribution provides a foundational understanding of the thematic spread and prevalence of discrimination discourse during societal events.

II. Social Network Analysis

To explore the structural dynamics of online communities, interaction data was visualized as social network graphs. Nodes represented individual users, while edges indicated interactions such as replies or retweets. The analysis involved:
- Community Detection: Identifying clusters of users and their interactions to uncover fragmented groups.
- Influence Mapping: Pinpointing key users driving the discourse within specific communities.
- Echo Chambers: Analyzing the cohesiveness of clusters and the extent of cross-group dialogue.

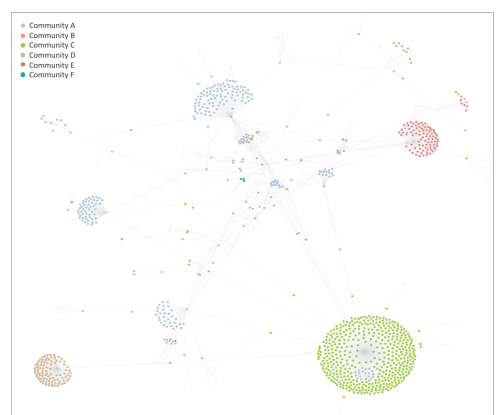

Fig. 1. Community-oriented Social Network

Figure 1 demonstrates the methodology for constructing community-oriented social networks, where distinct communities are represented as clusters differentiated by color. This visualization highlights the structural dynamics of user interactions, showcasing how communities form and dissolve over time in response to key events.

III. Temporal Analysis

To capture the dynamic nature of online interactions, a temporal analysis was conducted. This included:
- Event Correlation: Linking the peaks and troughs of discourse activity to specific real-world events.
- Topic Evolution: Tracking the growth, decline, and shifts in dominance of discourse fragments over time.

- Sentiment Shifts: Analyzing changes in public sentiment within different fragments of discrimination discourse.

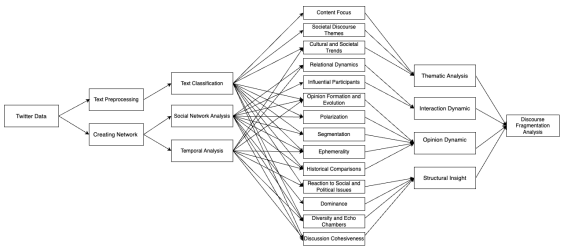

Fig. 2. Workflow of Discourse Fragmentation Analysis for Twitter Data

To outline the integration of thematic, structural, and temporal components, Figure 2 provides a comprehensive workflow for discourse fragmentation analysis. This framework ensures that all analytical layers interact cohesively to generate meaningful insights about discourse dynamics.

*E. Analytical Framework*

The study integrates a robust analytical framework encompassing thematic, structural, and temporal dimensions. The framework focuses on:
- Engagement Patterns: User interactions and their distribution across discrimination topics.
- Structural Cohesion: The density and connectivity of user communities within the discourse.
- Discourse Fragmentation: The formation of distinct clusters and their impact on overall conversation dynamics.

*F. Ethical Considerations*

The study adheres to ethical guidelines for handling publicly available social media data. User anonymity was ensured by excluding identifiable information, and all analyses were conducted in compliance with the terms of service of the platforms used. The focus remained on aggregate trends and interactions rather than individual behavior.

*G. Scope and Limitations*

This study is limited to specific social media platforms and events, which may not fully capture the diversity of online discourse. Additionally, while computational methods provide valuable insights into the structural and thematic aspects of fragmentation, they are less adept at capturing the nuanced context of individual interactions.

## IV. FINDINGS

This study investigates the fragmentation of online discourse, with a focus on the amplification of discrimination discourse and its structural and temporal dynamics. Rooted in Michael Sandel's philosophy, the findings highlight how social media platforms undermine community and shared reasoning by amplifying polarizing narratives, isolating users within ideological silos, and accelerating the volatility of digital conversations.

*A. Amplification of Discrimination Discourse*

The analysis reveals that discrimination discourse—encompassing sexism, racism, xenophobia, ableism, homophobia, and religious intolerance—gains significant traction in online spaces, particularly during moments of heightened political and social unrest. Real-world events act as catalysts for the emergence of discourse fragments, with topics rapidly shifting in response to external stimuli. During large-scale crises, such as the COVID-19 pandemic and the Black Lives Matter protests, discussions centered on discrimination intensified, reinforcing ideological divisions. The self-reinforcing nature of these discussions led to the dominance of certain discriminatory narratives within isolated digital communities, reinforcing ideological divisions and limiting collective reasoning [15].

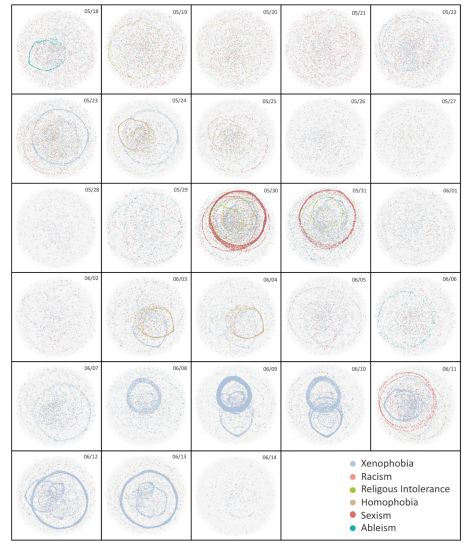

Fig. 3. San Francisco in the Black Lives Matter movement

Figure 3 illustrates how discussions on racial discrimination surged in response to the protests, demonstrating the dominance of race-related discourse in online interactions. The visual representation of fragmented discussions highlights the extent to which real-world events drive digital discourse, reinforcing Sandel's argument that community cohesion is weakened when ideological groups retreat into isolated conversations.

*B. Structural Fragmentation of Digital Communities*

The structural analysis of online interactions reveals tightly connected clusters, limiting cross-group dialogue and reinforcing echo chambers [16]. The fragmentation of discourse is not merely a function of individual preferences but is systematically reinforced by the architecture of social media platforms, where users engage predominantly with like-minded individuals. The presence of influential users or opinion leaders within these clusters further solidifies ideological divisions, as they amplify narratives that align with their audience's biases.

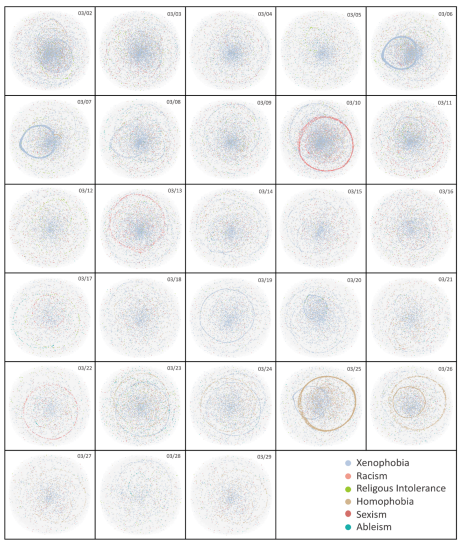

Fig. 4. Seattle during the COVID-19 pandemic

Figure 4 provides a structural visualization of how pandemic-related discussions coalesced into distinct clusters. This fragmentation reflects Sandel's concern that contemporary digital spaces hinder collective deliberation by fostering isolated moral communities rather than encouraging inclusive civic engagement. The figure illustrates how ideological segmentation manifests at a network level, making meaningful dialogue across ideological lines increasingly rare.

*C. Temporal Volatility of Discourse Fragments*

Discourse fragmentation is highly volatile, with dominant narratives emerging and dissipating rapidly due to algorithmic prioritization of trending content [17]. The lifespan of discrimination discourse is often transient, dictated by the social media attention cycle, in which trending topics gain temporary prominence before being replaced by new narratives. This pattern, driven by algorithmic amplification, prevents sustained engagement with critical societal issues. As a result, moral reasoning within digital spaces remains fragmented, preventing long-term collective reflection on discrimination and social justice.

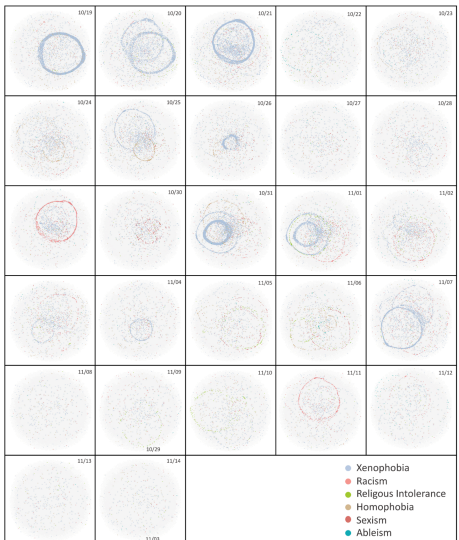

Fig. 5. San Francisco in the 2020 presidential race

Figure 5 illustrates the shifting nature of online discussions, showing how political events reshaped discourse dynamics. The visualization captures how different ideological groups engaged in discourse before, during, and after the election cycle, reflecting the instability of digital discussions. This volatility reinforces Sandel's critique of an individualized, market-driven society, where deliberative reasoning is subordinated to short-lived engagement with highly curated content.

*D. Role of Algorithms in Shaping Discourse*

The findings confirm that social media algorithms play an active role in amplifying ideological fragmentation. Content that evokes strong emotional responses—particularly outrage or moral indignation—tends to receive greater algorithmic visibility, reinforcing feedback loops that sustain polarized discourse. As users engage more with specific topics, they are disproportionately exposed to similar content, reducing their exposure to diverse perspectives and reinforcing ideological segmentation [18]. This process fundamentally undermines the principles of democratic deliberation and collective moral reasoning that Sandel envisions as essential to a just society.

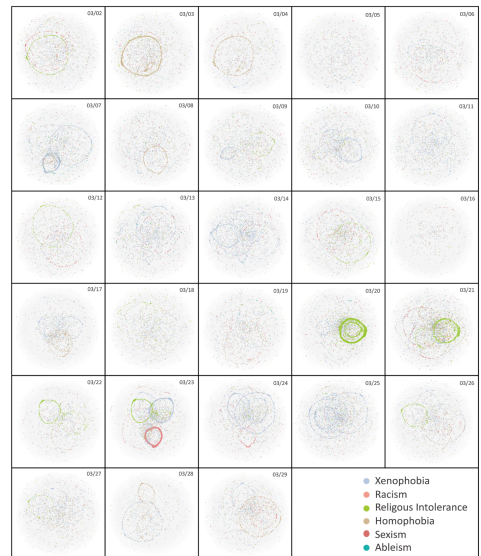

Fig. 6. New York City during the COVID-19 pandemic

Figure 6 highlights how algorithmic curation influenced discussions related to the pandemic. The figure demonstrates how certain narratives gained prominence over time, reinforcing existing biases rather than fostering inclusive discourse. This phenomenon aligns with previous research on algorithmic amplification, which suggests that platform design prioritizes engagement-driven content over balanced deliberation.

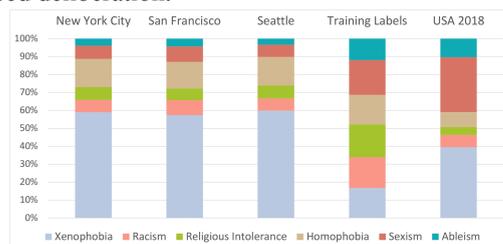

Fig. 7. Proportion of classes for each hate speech category

Figure 7 provides further evidence of algorithmic reinforcement by illustrating how users engaging in specific discriminatory discourse fragments were repeatedly exposed to similar narratives. This visual representation of content clustering highlights the cyclical nature of algorithmic amplification, which plays a crucial role in perpetuating fragmented discourse and limiting users' exposure to alternative perspectives. By creating insular communities that repeatedly engage with the same narratives, algorithms contribute to the erosion of the shared public sphere, reinforcing Sandel's warning that individualism, when left unchecked, weakens the foundations of civic discourse.

The findings of this study underscore the structural and thematic fragmentation of digital discourse, reinforcing Sandel's argument that a just society cannot exist without a shared moral framework. The amplification of discrimination discourse, the structural division of digital communities, and the ephemeral nature of online discussions collectively weaken the possibility of sustained public reasoning. Moreover, algorithmic prioritization of engagement-driven content exacerbates ideological segmentation, further reducing opportunities for collective moral engagement.

As Sandel warns, the erosion of community values in favor of hyper-individualized engagement ultimately leads to a more fragmented and polarized society. The findings confirm that digital platforms, by design, do not foster the kind of civic discourse necessary for democratic deliberation. Instead, they promote ideological silos, algorithmic reinforcement, and ephemeral discourse patterns that undermine the principles of collective reasoning and mutual accountability.

## V. Conclusions

The digital transformation of public discourse, as this study reveals, has not merely reshaped how we communicate but fundamentally altered the conditions necessary for democratic deliberation. Social media platforms, engineered to prioritize engagement through algorithmic amplification and filter bubbles, have fractured the public square into polarized echo chambers. These digital architectures amplify discriminatory narratives—from sexism and racism to religious intolerance—while eroding the shared moral framework that philosopher Michael Sandel argues is vital for a just society. As Sandel warns, "A just society can't be achieved simply by maximizing individual choice or by ensuring fair procedures; it depends on cultivating solidarity and shared responsibility". Yet, the hyper-individualized logic of social media, which mirrors the market-driven liberalism Sandel critiques, reduces civic discourse to a competition for attention, privileging reactivity over reflection and fragmentation over solidarity.

The findings of this research underscore a paradox: while digital platforms claim to democratize discourse, their algorithmic structures actively undermine the communal foundations of democracy. By fostering ideological silos and incentivizing divisive content, these platforms corrode the mutual accountability and collective reasoning Sandel deems essential. "When we decide that certain goods may be bought and sold," he writes, "we decide, at least implicitly, that it is appropriate to treat them as commodities". In this light, social media's commodification of discourse—trading deliberative depth for viral engagement—represents not merely a technical failure but a moral one, where the common good is subordinated to corporate and algorithmic interests.

Addressing this crisis demands more than technical fixes like algorithmic transparency or content moderation. It requires, as Sandel urges, "a politics of the common good" that recenters civic virtue and shared purpose in digital spaces. This entails reimagining platforms not as engines of individualized consumption but as infrastructures for fostering cross-ideological dialogue, nurturing empathy, and rebuilding collective agency. The challenge is both philosophical and practical: to resist the tyranny of algorithmic determinism and reclaim the digital public sphere as a space where solidarity, not polarization, defines the boundaries of discourse.

In the end, the fragmentation of online discourse is not inevitable—it is a choice embedded in the design and values of our technologies. To counter it, we must heed Sandel's call to "ask how we might govern ourselves better by deliberating together, in public, about how to value the social goods that define our common life". Only then can the digital age fulfill its promise of democratizing communication without sacrificing the shared understanding upon which justice depends.